\documentclass[11pt]{article}
\usepackage{longtable,supertabular}
\usepackage{amsmath}
\usepackage{amssymb}
\usepackage{latexsym}
\usepackage{cite}

\title{\bf Linear MSRD Codes with Various Matrix Sizes and Unrestricted Lengths}
\author{Hao Chen
  \thanks{Hao Chen is with the College of Information Science and Technology/Cyber Security, Jinan University, Guangzhou, Guangdong Province, 510632, China, haochen@jnu.edu.cn. The research of Hao Chen was supported by NSFC Grant 62032009.}}

\begin{document}

\maketitle
\begin{abstract}
A sum-rank-metric code attaining the Singleton bound is called maximum sum-rank distance (MSRD). MSRD codes have been constructed for some parameter cases. In this paper we construct a linear MSRD code over an arbitrary finite field ${\bf F}_q$ with various matrix sizes $n_1>n_2>\cdots>n_t$ satisfying $n_i \geq n_{i+1}^2+\cdots+n_t^2$ for $i=1, 2, \ldots, t-1$ for any given minimum sum-rank distance.\\

\end{abstract}

\section{Introduction}

Sum-rank-metric was first defined in \cite{NU} and there have been wide applications of sum-rank-metric codes in universal error correction and security in multishot network coding, space-time coding and coding for distributed storage, see \cite{NU,WZSS,SK,MK19,MK,MP1}. For definitions and fundamental properties of sum-rank-metric codes, we refer to \cite{MK,MP1,BGR,BGR1,CGLGMP}. Let $n_i \leq m_i$ be $t$ positive integers, $m_1 \geq m_2 \cdots \geq m_t$, $N=n_1+\cdots+n_t$.  Set $[n]=\{1, \ldots, n\}$.  Let  ${\bf F}_q^{n \times m}$ be the linear space of all $n \times m$ matrices over ${\bf F}_q$, and ${\bf F}_q^{(n_1, m_1), \ldots,(n_t, m_t)}={\bf F}_q^{n_1 \times m_1} \bigoplus \cdots \bigoplus {\bf F}_q^{n_t \times m_t}$ be the set of all ${\bf x}=({\bf x}_1,\ldots,{\bf x}_t)$, where ${\bf x}_i \in {\bf F}_q^{n_i \times m_i}$, $i=1,\ldots,t$.  Parameters $n_i \times m_i$, $i=1, \ldots, t$ are called matrix sizes of sum-rank-metric codes. We define the sum-rank weight of ${\bf x}$ $$wt_{sr}({\bf x}_1, \ldots, {\bf x}_t)=rank({\bf x}_1)+\cdots+rank({\bf x}_t),$$ and sum-rank distance $$d_{sr}({\bf x},{\bf y})=wt_{sr}({\bf x}-{\bf y}),$$ for ${\bf x}, {\bf y} \in {\bf F}_q^{(n_1,m_1), \ldots,(n_t,m_t)}$. This is a metric on ${\bf F}_q^{(n_1,m_1), \ldots,(n_t,m_t)}$. For a code ${\bf C} \subset {\bf F}_q^{(n_1,m_1), \ldots,(n_t,m_t)}$  its minimum sum-rank distance is defined by $$d_{sr}({\bf C})=\min_{{\bf x} \neq {\bf y}, {\bf x}, {\bf y} \in {\bf C}} d_{sr}({\bf x}-{\bf y}).$$ It is a basic goal of sum-rank-metric coding that to construct good sum-rank-metric codes with large cardinalities and large minimum sum-rank distances. For some basic upper bounds we refer to \cite{BGR} Section III and Section IV.\\

The sum-rank-metric is a generalization and combination of the Hamming metric and the rank-metric. When $t=1$, this is the rank-metric code. When $m_1=\cdots=m_t=m$ and $n_1=\cdots=n_t=n$, this is the $t$-sum-rank-metric code over ${\bf F}_{q^m}$ with the code length $N=nt$. When $n=1$, this is the Hamming error-correcting code case.\\

The Hamming weight $wt({\bf a})$ of a vector ${\bf a} \in {\bf F}_q^n$ is the number of non-zero coordinate positions. The Hamming distance $d_H({\bf a}, {\bf b})$ between two vectors ${\bf a}$ and ${\bf b}$ is the Hamming weight $wt({\bf a}-{\bf b})$. For a code ${\bf C} \subset {\bf F}_q^n$ the minimum Hamming distance is $$d_H({\bf C})=\min_{{\bf a} \neq {\bf b}} \{d_H({\bf a}, {\bf b}),  {\bf a} \in {\bf C}, {\bf b} \in {\bf C} \}.$$  For a linear $[n, k, d_H]_q$ code, the Singleton bound asserts $d_H \leq n-k+1$. When the equality holds, this code is an MDS code. We refer to \cite{HP} for the theory of Hamming error-correcting codes and numerous nice constructions. The main conjecture of MDS codes claims that the length of an MDS code over ${\bf F}_q$ is at most $q+1$, except some trivial exceptional cases. In \cite{Ball} the main conjecture of MDS codes was proved for codes over prime fields.\\

In this paper the repetition code in the Hamming metric ${\bf C}=\{(c_1, \ldots, c_n):c_1=\cdots=c_n\}$ over some finite field will be used. For each nonzero codeword in this code, the Hamming weight is exactly $n$.\\

We recall the rank-metric on the space ${\bf F}_q^{(m, n)}$ of size $m \times n$ matrices over ${\bf F}_q$, $d_r(A,B)= rank(A-B).$ For a rank-metric code ${\bf C} \subset {\bf F}_q^{(m, n)}$ its minimum rank-distance is $$d_r({\bf C})=\min_{A\neq B} \{d_r(A,B): A \in {\bf C}, B\in {\bf C} \}$$  The Singleton bound for a rank-metric code asserts that the number of codewords in ${\bf C}$ is upper bounded by $q^{\max\{m,n\}(\min\{m,n\}-d_r+1)}$ if the minimum rank distance of ${\bf C}$ is at least $d_r$, see \cite{Gabidulin}. \\

The Gabidulin code $Gab(n, v) \subset {\bf F}_q^{(n, n)}$ attains the Singleton bound. It is consisting of ${\bf F}_q$ linear mappings on ${\bf F}_q^n \cong {\bf F}_{q^n}$ defined by $q$-polynomials $a_0x+a_1x^q+\cdots+a_ix^{q^i}+\cdots+a_tx^{q^t}$, where $a_t,\ldots,a_0 \in {\bf F}_{q^n}$ are arbitrary elements in ${\bf F}_{q^n}$, see \cite{Gabidulin}. The rank-distance of is at least $n-t$ since there are at most $q^t$ roots in ${\bf F}_{q^n}$ for each such $q$-polynomial. There are  $q^{n(t+1)}$ such $q$-polynomials. Hence the size of the Gabidulin code is $q^{n(t+1)}$. This is an MRD code.  We refer to \cite{BNRS,Gola} for recent results on rank-metric codes and \cite{HTM2017,NHTRR2018} for recent results on MRD codes.\\

The Singleton upper bound for the rank-sum-metric was proved in \cite{MK,BGR}. The general form Theorem III.2 in \cite{BGR} is as follows. Let the minimum sum-rank distance $d$ can be written as the form $d_{sr}=\Sigma_{i=1}^{j-1} n_i+\delta+1$ where $0 \leq \delta \leq n_j-1$, then $$|{\bf C}| \leq q^{\Sigma_{i=j}^t n_im_i-m_j\delta}.$$ The code attaining this bound is called maximal sum-rank-metric distance (MSRD). When $m_1=\cdots=m_t=m$ this bound is of the form $$|{\bf C}| \leq q^{m(N-d_{sr}+1)},$$ see \cite{MP1}. It degenerates to the Singleton bound for the Hamming metric code when $m=1$ and degenerates to the Singleton bound for the rank-metric codes when $t=1$. We call the the difference $\Sigma_{i=j}^t m_in_i-m_j\delta-\log_q |{\bf C}|$ the defect of the sum-rank-metric code ${\bf C}$.\\

When $t \leq q-1$ and $N \leq q^m+1$ MSRD codes attaining the above Singleton bound were constructed in \cite{MP1}. They are called linearized Reed-Solomon codes, we also refer to \cite{Neri} for the further results. When $t=q$, it was proved in \cite{BGR} Example VI.9, MSRD codes may not exist. In \cite{BGR} the maximal block lengths of MSRD codes are upper bounded in Theorem VI. 12. In several other cases, for example, when the sum-metric minimum distance is $2$ or $N$, MSRD codes exist for all parameters, they were constructed in Section VII of \cite{BGR}. In \cite{MP20} more linear MSRD codes with the same matrix size defined over smaller fields were constructed by extended Moore matrices. Generalized sum-rank weights were defined for sum-rank-metric codes via optimal anticodes in \cite{CGLGMP}. It was proved in \cite{CGLGMP} that the generalized sum-rank weights of an MSRD code is determined by its block size, matrix size, dimension and distance parameters, as that of the generalized Hamming weights of an MDS codes. MSRD codes have applications on space-time coding, maximally recoverable LRC codes and partial-MDS codes, we refer to \cite{MK,CMST,SK}. \\

In \cite{MP21} Mart\'{i}nez-Pe\~{n}as proposed and studied sum-rank BCH codes of the matrix size $n_1=\cdots=n_t=n$, $m_1=\cdots=m_t=m$ as a generalization of Hamming BCH codes. These sum-rank-metric codes are linear over ${\bf F}_{q^m}$. The minimum sum-rank distance of a sum-rank BCH code is always bigger than or equal to its designed distance and the dimensions of these sum-rank BCH codes are lower bounded in \cite{MP21} Theorem 9. We refer to Tables V, VI and VII for many sum-rank BCH codes of parameters $n=m=2$, $q=2$.  In our recent preprint \cite{HCHEN}, numerous constructed sum-rank-metric codes are better than these sum-rank BCH codes. Some bounds and probabilistic constructions of sum-rank-metric codes were given in \cite{Puchinger,OPB}.\\

There has been very few constructed sum-rank-metric codes with various matrix sizes, except some examples of MSRD codes in \cite{BGR}. In previous papers \cite{MP1,MK19,MP21,BGR,BGR1,CGLGMP,Neri,NSZ21} about constructions of sum-rank-metric codes, most of them are concentrated in the matrix size case $n_1=n_2=\cdots=n_t, m_1=m_2=\cdots=m_t$.  In this paper we construct new linear MSRD codes over an arbitrary finite field ${\bf F}_q$ with the matrix sizes $n_1 \times n_1, n_2 \times n_2, \ldots, n_t\times n_t$, where $n_1 >n_2\cdots >n_t$ are $t$ positive integers satisfying $n_i \geq n_{i+1}^2+\cdots+n_t^2$, for $i=1, 2, \ldots, t-1$,  and any given minimum sum-rank distance $d_{sr}$. There is no restriction on the length of the code from the size $q$ of the finite field.  Our result illustrates that the theory of sum-rank-metric codes with various matrix sizes is basically different with the theory of sum-rank-codes with the same matrix size. This can be compared with the results obtained in \cite{MP1,BGR}. Moreover this is also quite different to the essence of the MDS main conjecture in the Hamming metric. On the other hand comparing with constructed non-trivial optimal LRC codes, quantum MDS codes and entanglement-assisted quantum MDS codes attaining the Singleton bound in \cite{GXY,CLZ,CZJL,Pellikaan}, code lengths are bounded by some $O(q^2)$. It is a surprise that for sum-rank-metric codes with various matrix sizes attaining the Singleton bound, the block length $t$ and the parameter $N=n_1+\cdots+n_t$ have no relation with the field ${\bf F}_q$.\\

\section{Block size two MSRD codes}

In this Section we first give linear MSRD codes with the block size $t=2$ for the convenience of understanding. First of all the matrix space ${\bf M}_{n \times n}({\bf F}_q)$ is identified with all $q$-polynomials $a_0 +a_1x^q+\cdots+a_{n-1}x^{q^{n-1}}$, where $a_0, a_1, \ldots, a_{n-1} \in {\bf F}_{q^n}$. \\

{\bf Theorem  2.1.} {\em Let $n_1 \geq n_2^2$ be two positive integers. Then a linear MSRD code ${\bf C}$ with the block size $2$, matrix sizes $n_1 \times n_1, n_2\times n_2$ over an arbitrary field ${\bf F}_q$ and any given minimum sum-rank distance can be constructed explicitly. }\\

{\bf Proof.} We discuss two cases. The 1st case is $d_{sr} \leq n_1-1$. Then the Singleton bound is $q^{n_1(n_1-d_{sr}+1)+n_2^2}$. The first part of ${\bf C}$ is consisting of $q^{n_1(n_1-d_{sr}+1)}$ codewords of the form $(a_0x+\cdots+a_{n_1-d_{sr}}x^{q^{n_1-d_{sr}}}, {\bf 0})$, where $a_0, a_1, \ldots, a_{n_1-d_{sr}}$ are $n-d_{sr}+1$ arbitrary elements on the field ${\bf F}_{q^{n_1}}$, and the $q$-polynomial is understood as a matrix in ${\bf F}_q^{(n_1, n_1)}$. the second matrix is zero matrix. It is clear that the first part is a linear code with the minimum sum-rank distance at least $n_1-(n_1-d_{sr})=d_{sr}$.\\

The second part corresponds to $q^{n_2^2}$ codewords. First of all we decompose the ${\bf F}_q$-linear space ${\bf F}_{q^{n_1}} \cdot x^{q^{n_1-d_{sr}+1}}$ as the direct sum $n_2$ linear subspaces ${\bf V}_1, \ldots, {\bf V}_{n_2}$,  of the dimension $\dim_{{\bf F}_q}({\bf V}_i)=n_2$, $i=1, \ldots, n_2$.. This is guaranteed from the condition $n_1 \geq n_2^2$. Then each of $n_2$ subparts of the second part is a dimension $n_2$ code consisting of all codewords of the form $(a_{n_1-d_{sr}+1}x^{q^{n_1-d_{sr}+1}}, b_i x^{q^i})$, where $a_{n-d_{sr}+1} \in {\bf V}_i$, $b_i \in {\bf F}_{q^{n_2}}$, and moreover they are the same after a suitable ${\bf F}_q$ linear space isomorphism of ${\bf V}_i$ to ${\bf F}_{q^{n_2}}$. Therefore the dimension of each subpart is $n_2$. The second part is the direct sum of all these $n_2$ subparts and it is obvious the dimension is $n_2^2$. The minimum sum-rank distance is at least $n_1-(n_1-d_{sr}+1)+1=d_{sr}$ since $a_{n_1-d_{sr}+1}$ and $b_i$ are nonzero for a nonzero codeword.\\

When $d_{sr}=n_1+d$, where $0 \leq d \leq n_2-1$. The Singleton bound is $q^{n_2(n_2-d+1)}$. The linear code ${\bf C}$ is the direct sum of $n_2-d+1$ subcodes of the dimension $n_2$. As in the first case we decompose the ${\bf F}_q$-linear space ${\bf F}_{q^{n_1}} \cdot x^{q^0}$ as the direct sum of $n_2$ linear subspaces ${\bf V}_1, \ldots, {\bf V}_{n_2}$,  of the dimension $\dim_{{\bf F}_q}({\bf V}_i)=n_2$, $i=1, \ldots, n_2$.. This is guaranteed from the condition $n_1 \geq n_2^2$. Then each of these $n_2-d+1$ linear subcodes is consisting of codewords of the form $(a_0^jx^{q^0}, b_j x^{q^{j-1}})$ where $a_0^j \in {\bf V}_j$ and $b_j \in {\bf F}_{q^{n_2}}$ for $j=1, \ldots, n_2-d+1$, and $a_0^j$ and $b_j$ are the same with a suitable linear isomorphism of ${\bf V}_j$ with ${\bf F}_{q^{n_2}}$. This is a dimension $n_2$ linear subcode. It is easy to verify that the minimum sum-rank distance is at least $n_1+n_2-(n_2-d)=n_1+d$. \\

It is easy to verify that all these linear subcodes are linear independent. Hence the conclusion is proved.\\

\section{MSRD code construction}

In this section we give our explicit constructions of MSRD codes with various matrix sizes. From the following result it is clear that the code cardinality attains the Singleton bound.\\

{\bf Theorem 3.1.} {\em Let $n_1>n_2>\cdots>n_t$ be $t$ positive integers. Let $d_{sr}=\Sigma_{i=1}^{j-1}n_i+d_1$ where $j \in \{1, \ldots, t\}$ and $0 \leq d_1 \leq n_j-1$ be the unique representation of the minimum sum-rank distance. Suppose that $n_1, \ldots, n_t$ and $d_{sr}$ satisfy \\
1) $n_{j-1} \geq n_j(n_j-d_1+1)+n_{j+1}^2+\cdots+n_t^2$;\\
2) $n_j \geq n_{j+1}^2+\cdots+n_t^2$.\\
Then a linear sum-rank-metric code with $q^{n_j(n_j-d_1+1)+\Sigma_{i=j+1}^t n_i^2}$ codewords over an arbitrary field ${\bf F}_q$ can be constructed explicitly.}\\

{\bf Proof.} In each block position we have linear independent ${\bf F}_q$-linear mappings $1, x^q, \ldots, x^{q^{n_i-1}}$ over ${\bf F}_{q^{n_i}}$ for $i=1, \ldots, t$.  In our construction many copies of repetition codes over ${\bf F}_{q^{n_i}}$, $i=j, j+1, \ldots, t$ is used. When each such repetition code is used, new $x^{q^v}$'s in some block position are introduced, so linear independence is guaranteed. It is important that coefficients of some $x^{q^{v_i}}$ at different block positions are not zero for a nonzero codeword.\\

For each $x^{q^v}$ at the $i$-th block  position, the set of all coefficients is the field ${\bf F}_{q^{n_i}}$. This is a ${\bf F}_q$-linear space of the dimension $n_i$, therefore the linear space of all coefficients of $x^{q^v}$ at the $i$-th position can be decomposed to the direct sum of $(n_j-d_1+1)+n_{j+1}+\cdots+n_t$ ${\bf F}_q$-linear subspaces ${\bf V}_w^i$, $i=1, 2, \ldots, j-1$ and $w=1, 2, \ldots, n_j-d_1+1+n_{j+1}+\cdots+n_t$, of dimensions $n_j, \ldots, n_j, n_{j+1}, \ldots, n_{j+1}, \ldots, n_t, \ldots, n_t$. If $i \leq j-1$, this is guaranteed from the condition $n_{j-1} \geq n_j(n_j-d_1+1)+n_{j+1}^2+\cdots+n_t^2$.\\

The dimension in the Singleton bound for sum-rank-metric codes is $n_j(n_j-d_1+1)+n_{j+1}^2+\cdots+n_t^2$. The first term $n_j(n_j-d_1+1)$ comes from $x^{q^0}, x^q, \ldots, x^{q^{n-d_1}}$ at the $j$-the block position. We use $n_j-d_1+1$ copies of length $j$ repetition code as in the proof of Theorem 2.1. Then for a nonzero codeword ${\bf c}=({\bf c}_1, \ldots, {\bf c}_j)$ in this repetition code, ${\bf c}_1, \ldots, {\bf c}_j$ are not zero. The dimension $n_j$ linear subspace of ${\bf F}_{q^{n_1}}$ of the coefficient $x^{q^0}$ at the 1st block position is used for ${\bf c}_1$ with a suitable base of ${\bf F}_{q^{n_1}}$, ......, the dimension $n_j$ linear subspace of ${\bf F}_{q^{n_{j-1}}}$ of the coefficient $x^{q^0}$ at the $j-1$-th block position is used for ${\bf c}_{j-1}$ with a suitable base of ${\bf F}_{q^{n_{j-1}}}$, at the $j$-th position, the coefficient $x^{q^0}, \ldots, x^{q^{n_j-d_1}}$ at the $j$-th block position is used for ${\bf c}_j$, for the $j+1, \ldots, t$-th block positions, zero $q$-polynomials are used. Then we have $q^{n_j(n_j-d_1+1)}$ codewords in the constructed sum-rank-metric code from these $n_j-d_1+1$ copies of the repetition code. The minimum sum-rank distance is at least $n_1+\cdots+n_{j-1}+(n_j-(n_j-d_1))=n_1+\cdots+n_{j-1}+d_1$, since ${\bf c}_1, \ldots, {\bf c}_j$ are not zero.\\

For the $(i+1)$-th term $n_{j+i}^2$ in the dimension of the Singleton bound for sum-rank-metric codes, we use $n_{j+i}$ copies of the length $j+1)$ repetition code. For a codeword ${\bf c}=({\bf c}_1, \ldots, {\bf c}_{j+1})$, it is obvious that ${\bf c}_1, \ldots, {\bf c}_{j+1}$ are not zero for a nonzero codeword. The dimension $n_{j+i}$ linear subspace of ${\bf F}_{q^{n_1}}$ of the coefficient $x^{q^0}$ at the 1st block position is used for ${\bf c}_1$ with a suitable base of ${\bf F}_{q^{n_1}}$, ......, the dimension $n_{j+i}$ linear subspace of ${\bf F}_{q^{n_{j-1}}}$ of the coefficient $x^{q^0}$ at the $j-1$-th block position is used for ${\bf c}_{j-1}$ with a suitable base of ${\bf F}_{q^{n_{j-1}}}$, at the $j$-th position, the dimension $n_{j+i}$ linear subspace of ${\bf F}_{q^{n_j}}$ of the coefficient of $x^{q^{n_j-d_1+1}}$ is used for ${\bf c}_j$ with a suitable base since $n_j \geq n_{j+1}^2+\cdots+n_t^2$, at the $j+1, \ldots, j+i-1$-th positions, only zero $q$-polynomials are used, the coefficient $x^{q^0}, \ldots, x^{q^{n_{j+i}-1}}$ at the $j+i$-th block position is used for ${\bf c}_{j+1}$, for the $j+i+1, \ldots, t$-th block positions,  zero $q$-polynomials are used. Then we have $q^{n_{j+1}^2}$ codewords in the constructed sum-rank-metric code from these $n_{j+1}$ copies of the repetition code. The minimum sum-rank distance is at least $n_1+\cdots+n_{j-1}+(n_j-(n_j-d_1+1))+1=n_1+\cdots+n_{j-1}+d_1$, since ${\bf c}_1, \ldots, {\bf c}_{j+i}$ are not zero.\\

The above constructed sum-rank-metric code is ${\bf F}_q$-linear. For two different codewords ${\bf x}_1, {\bf x}_2$ in the above constructed sum-rank-metric code, the difference ${\bf x}_1-{\bf x}_2$ has the rank at least $n_i$ at the $i$-th block position, for $i=1, \ldots, t-1$, and the sum-rank at least $d_1$ at $j, j+1, \ldots t$-th block positions. Then the minimum sum-rank distance is at least $n_1+\cdots+n_{j-1}+d_1$. The conclusion is proved.\\

{\bf Corollary 3.1.} {\em Let $n_1 >n_2 > \cdots >n_t$ be $t$ positive integers satisfying $n_i \geq n_{i+1}^2+\cdots+n_t^2$ for $i=1, 2, \ldots, t-1$. Then a linear MSRD code over an arbitrary finite field ${\bf F}_q$ with any given minimum sum-rank distance can be constructed explicitly.}\\

\section{Conclusion}
In this paper explicit inear MSRD codes with matrix sizes $n_1 \times n_1, \ldots, n_t \times n_t$ satisfying $n_i \geq n_{i+1}^2+\cdots+n_t^2$, $i=1, 2, \ldots, t-1$, over an arbitrary finite field,  are constructed for all possible minimum sum-rank distances.  Comparing with previous constructions of MSRD codes with the same matrix size, our results show that linear sum-rank-metric codes with various matrix sizes are quite different. \\

\end{document}